\documentclass[amssymb,aps,twocolumn,showpacs,nofootinbib]{revtex4}
\usepackage[]{graphicx}
\usepackage{amsmath,amssymb}

\begin{document}

\title{QUANTUM KEY DISTRIBUTION FROM A RANDOM SEED}

\author{Eduin H. SERNA}\email{ehs@adaptun.com}
\affiliation{Quantum Technology Laboratory, ADAPTUN SAS,\\ Carrera 69A No. 44A-36 Torre 3 Int 502, Medell\'in, Colombia}

\begin{abstract}
It is designed a new quantum cryptography protocol that generates various secret and secure keys of the same size of the transmitted qubits, implying zero information losses between the interlocutors. Besides, generates key swapping between the two recipients of photons, without even sharing a past between them. This protocol differs from BB84 just in the classic procedures, using a random seed and asymmetric cryptography.    

\end{abstract}


\maketitle

\section{Introduction }

One of the most difficult practical problems when conducting secure communication is the key distribution. Shannon in 1949 \cite{S49}, established that if the key, is random, has the same length of the message to code and is used one time only, then it is guaranteed statistical independence of the cryptogram with respect to the message.  In fact, it is the only system demonstrated to be mathematically secure. However, the need to distribute and store securely the keys,  generally long and just for one use, limits its implementation.  This panorama remained the same until 1984, year in which Bennett and Brassard presented the first quantum key distribution protocol (QKD), called BB84 \cite{BB84}. This protocol is that the interlocutors (Alice, Bob) transmit the key through a quantum channel. The quantum contribution to the security of the process is that an eavesdropper cannot extract information without revealing its presence to the interlocutors.

Essentially the QKD protocols BB84 , B92 \cite{B92}, E91 \cite{E91} and SARG04 \cite{SARG04} apply two processes: - \textit{Raw key exchange}, obtained from the initial interpretation of the quantum states exchanged between Alice and Bob. - \textit{Public Reconciliation}, where, Alice and Bob obtain the secret key from the published information.  Thus, QKD is the only method physically secure to exchange secret keys \cite{G02}. 
\\

In this article it is presented a new protocol that is identical to the BB84 for all the quantum manipulations, but differs from it by using \textit{Private Reconciliation} from a \textit{Random Seed} and \textit{Asymmetric Cryptography}. Thus allowing the generation of larger secure keys.

This article is designed as follows: Sec II presents the formalism of the BB84 protocol, Sec III present the new protocol and its differences from the BB84 and Sec IV contains the conclusions.

\section{BB84 Protocol}

\subsection{Quantum Part}

Alice and Bob exchange a set of encoded photons according to four states $|0\rangle,|1\rangle,|+\rangle,|-\rangle,$  which gather forming two basis with orthogonal states  $\mathfrak{B}_0=\{|0\rangle,|1\rangle\}$ and $\mathfrak{B}_1=\{|+\rangle,|-\rangle\},$ where $|\pm\rangle=\frac{1}{\sqrt{2}}\left( |0\rangle\pm|1\rangle\right)$.Coding the binary value $0$ to the states  $|0\rangle$ and $|+\rangle,$  and the binary value $1$ to the states  $|1\rangle$ and $|-\rangle.$ For simplicity it is denoted  $|\psi_{00}\rangle\equiv|0\rangle,$ $|\psi_{01}\rangle\equiv|1\rangle,$ $|\psi_{10}\rangle\equiv|+\rangle,$ $|\psi_{11}\rangle\equiv|-\rangle.$ 
\\

\begin{itemize}
\item\textit{Raw Key Exchange }
	\begin{itemize}
	\item Alice generates two random strings with the same length, $N$, The strings will correspond to the keys Alice wants to share with Bob, one of binary basis $s_1s_2\dots s_N$ and the other of binary values $i_1i_2\dots i_N.$ From the elements occupying the concrete position, $k$, in both strings, Alice obtains the associated state $|\psi_{s_ki_k}\rangle$ and sends it to Bob through a quantum channel.
	\item Bob generates a random string of binary basis $m_1m_2\dots m_N$ that will correspond to the key it wants to share with Alice. Bob measures each received state $|\psi_{s_ki_k}\rangle$ in the corresponding base  $\mathfrak{B}_{m_k}$, obtaining a binary string  $a_1a_2...a_N$.
	\end{itemize}	
\end{itemize}	

	\subsection{Classic Part}

Alice and Bob exchange a set of binary strings.
\begin{itemize}
\item \textit{Public Reconciliation}
	\begin{itemize}
	\item Bob sends to Alice the sequence of basis  $m_1m_2\dots m_N,$ through an public channel authenticated.
	\item Alice compares the strings $s_1s_2\dots s_N$ and $m_1m_2\dots m_N,$ sending to Bob the binary sequence $l_1l_2\dots l_N$ with $l_k=s_k\oplus m_k.$

	\item Alice and Bob share now a sequence of binary values  $i_k=a_k$ formed by $l_k=0$.
	\end{itemize}
	
\end{itemize}

An example of the BB84 is given in the Table 1. In perfect conditions Alice and Bob share and generate an identical random key.

\section{New Protocol}
\subsection{Quantum Part}
\begin{itemize}
\item \textit{Raw Key Exchange - BB84}
\item \textit{Random seed }

\begin{itemize}
\item Alice or Bob publish a random binary string $x_1x_2\dots x_N.$ 
\end{itemize}	

\item \textit{Missing Key Exchange}
\begin{itemize}
\item Alice sums  $s_k\oplus x_k,$ $k=1,2,...,N.$ Obtaining a sequence of binary basis  $t_1t_2\dots t_N$ and generates other random string of binary values $j_1j_2\dots j_N$ that will correspond  to other key  that it wants to exchange with Bob. From the elements occupying a concrete position, $k$, of the preceding strings,  Alice obtains the associated state $|\psi_{t_kj_k}\rangle$ and sends it to Bob through a quantum channel.

\item Bob sums $(1\oplus m_k)\oplus x_k,$ $k=1,2,...,N.$ Obtaining the string of binary basis $n_1n_2\dots n_N$ and measures each received state $|\psi_{t_kj_k}\rangle$ with the corresponding base $\mathfrak{B}_{n_k}$ generating the string $b_1b_2\dots b_N.$
\end{itemize}
\end{itemize}
\subsection{Classic Part}
Alice and Bob exchange a set of binary strings and apply in different binary arrangements the function $f$ defined as follows:
$$f(z,x,y):=\left \{ \begin{matrix} x,&z=0 \\ 
																		y,&z=1
							       \end{matrix}
						\right.
$$
\begin{table}[htbp]
    \begin{tabular}{|l|cccccccc|}
    \hline
1a: $s\equiv\textit{key}_s$ & 1 & 0 & 1 & 0 & 0 & 0 & 0 & 0 \\ \hline		
2a: $i\equiv\textit{key}_i$ & 0 & 1 & 1 & 0 & 1 & 1 & 0 & 0 \\ \hline    
3a: $|\psi_{si}\rangle$ & $|\psi_{10}\rangle$ & $|\psi_{01}\rangle$ & $|\psi_{11}\rangle$ & $|\psi_{00}\rangle$ & $|\psi_{01}\rangle$ & $|\psi_{01}\rangle$ & $|\psi_{00}\rangle$ & $|\psi_{00}\rangle$ \\ \hline \hline		
1b: $m\equiv\textit{key}_m$ & 0 & 1 & 1 & 0 & 0 & 1 & 1 & 0 \\ \hline
2b: $\mathfrak{B}_{m}$ & $\mathfrak{B}_{0}$ & $\mathfrak{B}_{1}$ & $\mathfrak{B}_{1}$ & $\mathfrak{B}_{0}$ & $\mathfrak{B}_{0}$ & $\mathfrak{B}_{1}$ & $\mathfrak{B}_{1}$ & $\mathfrak{B}_{0}$ \\ \hline
3b: $a$ & $\ast$ & $\ast$ & 1 & 0 & 1 & $\ast$ & $\ast$ & 0 \\ \hline\hline
4a BB84: $l$ & 1 & 1 & 0 & 0 & 0 & 1 & 1 & 0 \\ \hline
BB84: $\textit{key}$&  &  & 1 & 0 & 1 &  &  & 0 \\ \hline\hline		
4ab: $x$ & 1 & 1 & 0 & 0 & 0 & 1 & 1 & 0 \\ \hline	\hline			
5a: $t$ & 0 & 1 & 1 & 0 & 0 & 1 & 1 & 0 \\ \hline
6a: $j\equiv\textit{key}_j$ & 1 & 0 & 1 & 0 & 1 & 0 & 0 & 1 \\ \hline		
7a: $|\psi_{tj}\rangle$ & $|\psi_{01}\rangle$ & $|\psi_{10}\rangle$ & $|\psi_{11}\rangle$ & $|\psi_{00}\rangle$ & $|\psi_{01}\rangle$ & $|\psi_{10}\rangle$ & $|\psi_{10}\rangle$ & $|\psi_{01}\rangle$ \\ \hline \hline		
5b: $n$ & 0 & 1 & 0 & 1 & 1 & 1 & 1 & 1 \\ \hline
6b: $\mathfrak{B}_{n}$ & $\mathfrak{B}_{0}$ & $\mathfrak{B}_{1}$ & $\mathfrak{B}_{0}$ & $\mathfrak{B}_{1}$ & $\mathfrak{B}_{1}$ & $\mathfrak{B}_{1}$ & $\mathfrak{B}_{1}$ & $\mathfrak{B}_{1}$  \\ \hline		
7b: $b$ & 1 & 0 & $\ast$ & $\ast$ & $\ast$ & 0 & 0 & $\ast$ \\ \hline \hline		
8a: $y$ & 1 & 1 & 0 & 0 & 0 & 1 & 0 & 1 \\ \hline\hline		
8b: $u$ & $\ast$ & 0 & $\ast$ & 1 & 0 & 0 & 1 & 1 \\ \hline
9b: $v$ & 1 & $\ast$ & 1 & $\ast$ & $\ast$ & $\ast$ & $\ast$ & $\ast$ \\ \hline\hline
9a: $\textit{key}_m$ & 0 & 1 & 1 & 0 & 0 & 1 & 1 & 0 \\ \hline
10a: $l$ & 1 & 1 & 0 & 0 & 0 & 1 & 1 & 0 \\ \hline\hline 
10b: $\textit{key}_s$ & 1 & 0 & 1 & 0 & 0 & 0 & 0 & 0 \\ \hline		
11b: $\textit{key}_i$ & 0 & 1 & 1 & 0 & 1 & 1 & 0 & 0 \\ \hline
12b: $\textit{key}_j$ & 1 & 0 & 1 & 0 & 1 & 0 & 0 & 1 \\ \hline
    \end{tabular}
		\caption{The Proposed Protocol from Alice (a) to Bob (b), $N=8.$ The steps 1a-3a and 1b-3b are equal to BB84. Where $\ast\in\{0,1\},$
		$t=s\oplus x,$ $n=1\oplus x\oplus m,$ $y=i\oplus j,$	$u=n\oplus f(m,a,b\oplus y),$
		$v=n\oplus f(m,b,a\oplus y),$	$\textit{key}_m\equiv t\oplus f(s,(1\oplus i)\oplus u,j\oplus v),$ $l=s\oplus m,$ $\textit{key}_s\equiv m\oplus l,$ $\textit{key}_i\equiv f(l,a,b\oplus y),$	y $\textit{key}_j\equiv f(l,a\oplus y,b).$}		
\end{table}
\begin{itemize}
\item \textit{Asymmetric Cryptography}
\begin{itemize}
\item Alice sums  $i_k\oplus j_k,$ $k=1,2,...,N.$ Obtaining the binary string  $y_1y_2\dots y_N$ that sends to Bob. 
\item Bob encrypt  $m_k$ in $u_k$ and $v_k$ with 
$$u_k=n_k\oplus f(m_k,a_k,b_k\oplus y_k), 
$$
$$
v_k=n_k\oplus f(m_k,b_k,a_k\oplus y_k).
$$
$k=1,2,...,N.$ Obtaining the public strings $u_1u_2\dots u_N$ y $v_1v_2\dots v_N$ that sends to Alice.
\item Alice sums
$$t_k\oplus f(s_k,(1\oplus i_k)\oplus u_k,j_k\oplus v_k)$$
Decrypting $m_k$ for $k=1,2,...,N.$ Obtaining the private string  $m_1m_2\ldots m_N$ of Bob.
\end{itemize}
\end{itemize}

\begin{itemize}
\item \textit{Private Reconciliation}
	\begin{itemize}
	\item Alice compares the strings $s_1s_2\dots s_N$ and $m_1m_2\dots m_N,$ sending to Bob the binary sequence $l_1l_2\dots l_N$ with $l_k=s_k\oplus m_k.$
	\item Bob sums  $m_k\oplus l_k,$ $k=1,2,...,N.$ Obtaining the private string $s_1s_2\dots s_N$ of Alice and apply:
		$$
		f(l_k,a_k,b_k\oplus y_k)\equiv i_k
		$$
		$$
		f(l_k,a_k\oplus y_k,b_k)\equiv j_k,
		$$
	$k=1,2,...,N.$ Obtaining the private strings of  Alice $i_1i_2\dots i_N,$ y $j_1j_2\dots j_N.$
	\end{itemize}
An example of this protocol is given in Table 1. In perfect conditions Alice and Bob share four secret and secure keys with length  $N,$ $m_1m_2\dots m_N,$ $s_1s_2\dots s_N,$ $i_1i_2\dots i_N,$ y $j_1j_2\dots j_N.$
\end{itemize}

\subsection{Quantum Key Swapping} \label{sw}
If Central is an emitting source of photons of BB84, B92 or E91, Alice and Bob are recipients of these photons. And following the steps of the previous protocol, both generate a common key without even sharing a past between them, due that Central compares the sequences $m_{Alice}$ and $m_{Bob}$ in the \textit{Private Reconciliation} and informs in which they coincided. Besides, If Central used the same encoded states in \textit{Raw and Missing key Exchange}  for Alice and Bob, they share the private strings of Central.

\section{CONCLUSIONS} 

In summary, It has been demonstrated that using a \textit{random seed} over a set of photons and \textit{asymmetric cryptograph}y over the encoded bits, the QKD becomes a process of zero information losses, where the percentage of coincidence of the reconciliated key against the size of the raw key is 100\% unlike the BB84 in which the expected is 50\%. Besides, this protocol as the SARG04 is identical to the BB84 for all the quantum manipulations and differs only in the classic procedure. Thus, this protocol can be implemented in existing devices without modifications.

\end{document}